\documentclass
      [aip,jcp,preprint,amsmath,amssymb,superscriptaddress,groupedaddress]{revtex4-1}

\usepackage{graphicx} 
\usepackage{longtable}

\def\bea{\begin{eqnarray}}
\def\eea{\end{eqnarray}}
\def\ben{\begin{equation}}
\def\een{\end{equation}}
\def\benu{\begin{enumerate}}
	\def\enu{\end{enumerate}}
\def\beal{\begin{equation}\begin{aligned}} 
\def\eeal{\end{aligned}\end{equation}}

\def\n{n}

\def\sss{\scriptscriptstyle\rm}

\def\br{{\mathbf r}}
\def\bR{{\mathbf R}}

\def\d3{{\rm d^3}}
\def\intdhr{\int\d3{r}}
\def\intdhrs{\intdhr\,}

\def\s{_{\sss S}}
\def\p{_{\sss p}}
\def\f{_{\sss f}}

\def\nad{^{\rm nad}}

\def\unpol{^{\rm unpol}}

\usepackage{xspace}

\def\Tsnad{T\s\nad[\{n_\alpha\}]}
\def\epkin{t\s\nad}

\def\Qf{Q_f(\br)}

\begin{document}

\title{Constructing a Non-additive Non-interacting Kinetic Energy Functional Approximation for Covalent Bonds from Exact Conditions}

\author{Kaili Jiang}
\affiliation{Department of Physics and Astronomy, Purdue University, 525 Northwestern Ave., West Lafayette, IN 47907, USA}

\author{Jonathan Nafziger}
\affiliation{Department of Chemistry, Purdue University, 560 Oval Dr., West Lafayette IN 47907, USA}

\author{Adam Wasserman}
\email[Corresponding Author: ]{awasser@purdue.edu}
\affiliation{Department of Chemistry, Purdue University, 560 Oval Dr., West Lafayette IN 47907, USA}
\affiliation{Department of Physics and Astronomy, Purdue University, 525 Northwestern Ave., West Lafayette, IN 47907, USA}


%
\begin{abstract}
We present a non-decomposable approximation for the non-additive non-interacting kinetic energy (NAKE) for covalent bonds based on the exact behavior of the von Weizs\"{a}cker (vW) functional in regions dominated by one orbital.
This {\em covalent approximation} (CA) seamlessly combines the vW and the Thomas-Fermi (TF) functional with a switching function of the fragment densities constructed to satisfy exact constraints. 
It also makes use of ensembles and fractionally-occupied spin-orbitals to yield highly accurate NAKE for stretched bonds while outperforming other standard NAKE approximations near equilibrium bond lengths. 
We tested the CA within Partition-Density Functional Theory (P-DFT) and demonstrated its potential to enable fast and accurate P-DFT calculations. 

\end{abstract}

\maketitle

\section{Introduction}

Partition Density Functional Theory (P-DFT) \cite{CW07,EBCW10,NW14} is a quantum-embedding method \cite{SC16} that has achieved some success in solving the delocalization and static-correlation errors \cite{CMY08a} of approximate DFT calculations\cite{NW15}.
In P-DFT, a system is partitioned into non-interacting fragments and a unique, global partition potential is introduced to compensate for the fragment-fragment interactions. 
The total energy equals the sum of the fragment energy plus an interacting term called partition energy $E\p$.
One challenge of P-DFT is to efficiently obtain accurate approximations to the kinetic component of $E\p$, the non-additive non-interacting kinetic energy (NAKE), defined by:
\begin{equation}\label{eqn:Tsnad}
\Tsnad = T\s[n]-\sum_{a}T\s[n_\alpha],
\end{equation}
where $\Tsnad$ is the NAKE as a functional of the set of fragment densities $\{n_\alpha(\br)\}$, and $T\s[n]$ and $T\s[n_\alpha]$ are the non-interacting kinetic energy (KE) for the whole system and for fragment $\alpha$, respectively.
In most cases, the exact form of the NAKE as an explicit functional of $\{n_\alpha(\br)\}$ is unknown.

One general approach to obtain the {\em exact} NAKE for a given choice of approximate exchange-correlation (XC) functional is to evaluate it as an implicit functional of the set of fragment densities. 
This approach typically involves computationally expensive inversion methods. \cite{FJN+10,GAMMI10,HPC11,NWW11,NJW17,JW17} 
The NAKE can also be obtained much more efficiently via explicit functional approximations.
There are two categories of NAKE approximations: {\em decomposable} and {\em non-decomposable} approximations\cite{JN14}.
Decomposable approximations are constructed by applying known approximations for the non-interacting KE into Eq. \ref{eqn:Tsnad}.
Many approximate expressions of the non-interacting KE\cite{Tho26,Fer27,Wei35,KP56,Kir57a,Gol57,YT65,Bal72,Lie81,TW02a,CFLDS11,LFCDS11,LLP91,FR95,LC94,Tha92} have been proposed for use in orbital-free DFT (OF-DFT)\cite{WC02} calculations.
However, these approximations lack the accuracy and transferability that is needed to treat not only the non-interacting KE for systems in diverse scenarios\cite{WC02},
but also the NAKE\cite{NJW17,JNW18}.
This becomes a real issue when applying approximations of the NAKE with strongly interacting systems involving covalent bonds, as the NAKE in these systems is comparable to other components of the ground-state energy. 
On the other hand, non-decomposable approximations of the NAKE, which do not require approximating the full $T\s[n]$, have shown to be a viable approach for weakly interacting fragments, at least in the case of rare-gas dimers \cite{JNW18}.

Another challenge with P-DFT calculations of covalent bonds is that they typically involve fractional charges and spins. 
There are two different treatments to handle non-integer charges and spins: ensemble treatments (ENS) and fractionally occupied orbitals (FOO). \cite{NJW17} 
In the case of ENS, the energy functional is evaluated with a set of ensemble components of spin densities with integer numbers of electrons, whereas in the case of FOO, the highest occupied molecular orbital is considered to be fractionally occupied and the energy functional is evaluated with spin densities having a non-integer number of electrons.
When the exact NAKE and XC functionals are used, both treatments yield the same energy as long as the total density is the same. \cite{CMY08}
However, this is not the case when an approximate NAKE is employed.
For example, for several covalent dimers with approximated NAKE, ENS consistently yields more accurate NAKEs than FOO at equilibrium separations but develops large static-correlation errors for stretched bonds. With the FOO treatment of fractional charges, the NAKE correctly vanishes when stretching bonds. \cite{NJW17}

In this work, we study the dissociation behavior of alkali dimers. 
Their binding region is dominated by one orbital in which the NAKE can be accurately approximated by the von Weizs\"{a}cker (vW)\cite{Wei35} functional.
We thus propose a {\em covalent approximation} to the NAKE, a non-decomposable expression in which a simple switching functional of the fragment densities is used to turn on and off the Thomas-Fermi (TF)\cite{Tho26,Fer27} and vW functionals as the one-orbital limit is approached. 
Based on our analysis of the NAKE through the lens of ENS and FOO treatments (Sec. \ref{sec:Tsnad}) and on the observed behavior of the NAKE per particle (Sec. \ref{sec:behav_ep_kin}), we explain the derivation of our covalent approximation in Sec. \ref{sec:CA} and discuss the results for alkali dimers and a few other covalent dimers in Sec. \ref{sec:results}.



\section{NAKE}\label{sec:Tsnad}

A useful tool to study the behavior of the NAKE in different regions is the NAKE per particle $\epkin$, which is defined by any functional that satisfies
\begin{equation}\label{eqn:def_tsnad}
\Tsnad = \intdhrs n(\br)\epkin[\{\n_\alpha\}](\br)
\end{equation}
It can be calculated using the following equation:
\begin{equation}\label{eqn:tsnad}
\epkin[\{\n_\alpha\}](\br) = t\s[n](\br)-\sum_{\alpha}\frac{n_\alpha(\br)}{n(\br)}t\s[n_\alpha](\br)
\end{equation}
where $t\s[n](\br)$ and $t\s[n_\alpha](\br)$ are the non-interacting kinetic energy per particle for the whole system and for fragments, respectively.
Although the NAKE per particle defined by Eq. \ref{eqn:def_tsnad} is ambiguous because any term that integrates to 0 over the whole space can be added to a valid $\epkin$ to create another valid $\epkin$, this ambiguity can be removed when $\epkin$ is calculated through Eq. \ref{eqn:tsnad} by enforcing the {\em same} form of $t\s$ to be used for the whole system and for the fragments.

For spin-polarized systems, $n_\alpha=n_{\alpha\uparrow}+n_{\alpha\downarrow}$, we have \cite{Bur07}
\begin{equation}\label{eqn:Ts_pol}
T\s[n_{\alpha\uparrow},n_{\alpha\downarrow}]=\frac{1}{2}(T\s\unpol[2n_{\alpha\uparrow}]+T\s\unpol[2n_{\alpha\downarrow}])
\end{equation}
and
\begin{equation}\label{eqn:ts_pol}
t\s[n_{\alpha\uparrow},n_{\alpha\downarrow}](\br)=\frac{1}{2}(t\s\unpol[2n_{\alpha\uparrow}](\br)+t\s\unpol[2n_{\alpha\downarrow}](\br))
\end{equation}

As mentioned in the introduction, fractional charges and spins can be treated with ENS or FOO. 
A detailed comparison of these two methods in P-DFT can be found in ref. \cite{NJW17}. 
Here is a brief summary:

With ENS, the KE per particle for fragment $\alpha$ is written in the general form
\begin{equation}\label{eqn:ts_ens}
t\s^\mathrm{ENS}[n_{\alpha}](\br)=\sum_{i}f_{i\alpha}t\s[n_{i\alpha\uparrow},n_{i\alpha\downarrow}](\br)
\end{equation}
where $i$ is the ensemble component index and $f_{i\alpha}$ is the ensemble coefficient.
Substituting Eq. \ref{eqn:ts_ens} into Eq. \ref{eqn:ts_pol}, we have
\begin{equation}\label{eqn:ts_ens2}
t\s^\mathrm{ENS}[n_{\alpha}](\br)=\frac{1}{2}\sum_{i,\sigma}f_{i\alpha}t\s\unpol[2n_{i\alpha\sigma}](\br)
\end{equation}
On the other hand, with FOO, we have
\begin{equation}\label{eqn:ts_foo}
t\s^\mathrm{FOO}[n_{\alpha}](\br)=\frac{1}{2}\sum_{\sigma}t\s\unpol[2n_{\alpha\sigma}](\br)
\end{equation}
where with the same fragment density, $n_{i\alpha\sigma}$ and $n_{\alpha\sigma}$ have the following relationship:
\begin{equation}\label{eqn:density_conv}
n_{\alpha\sigma}(\br) = \sum_{i}f_{i\alpha}n_{i\alpha\sigma}(\br)
\end{equation}

As mentioned in the introduction, approximate ENS-NAKEs have large static-correlation errors, but FOO-NAKEs do not.
This is because approximate kinetic-energy functionals fail to reproduce the exact behavior that the NAKE should scale linearly with the density when the total number of electrons increases from $N$ to $N+1$, where $N$ is any non-negative integer \cite{NW14}.
For covalent dimers, the ENS-NAKE per particle for a stretched configuration develops incorrect features in the core regions, but the FOO-NAKEs remain vanishingly small there, as shown numerically in Sec. \ref{sec:behav_ep_kin}.
Here we provide a mathematical explanation for this behavior:

Consider a diatomic system partitioned into two fragments labeled by $L$ and $R$. 
Under the FOO treatment, the NAKE per particle can be written as
\begin{eqnarray}
t\s^\mathrm{nad,FOO}[n_L,n_R](\br)&=&\frac{1}{2}\sum_{\sigma}t\s\unpol[2(n_{L\sigma}+n_{R\sigma})](\br)\nonumber\\
& &-\frac{1}{2}\sum_{\sigma}t\s\unpol[2n_{L\sigma}](\br)-\frac{1}{2}\sum_{\sigma}t\s\unpol[2n_{R\sigma}](\br).
\end{eqnarray}
The fragment densities typically have exponential asymptotic behavior \cite{NW14}, which means that at large inter-nuclear separations $n_R\ll n_L$ and $\nabla n_R\ll\nabla n_L$ in the core region of the left nucleus. 
With a local (LDA) or semi-local (GGA) approximation, the influence of the right fragment density on the molecular $t\s$ evaluated close to left fragment can be ignored:
\begin{equation}
t\s\unpol[2(n_{L\sigma}+n_{R\sigma})](\br\rightarrow\bR_L)\approx t\s\unpol[2n_{L\sigma}](\br\rightarrow\bR_L).
\end{equation}
The $t\s$ of the right fragment in the left region can also be ignored:
\begin{equation}
t\s\unpol[2n_{R\sigma}](\br\rightarrow\bR_L)\approx 0.
\end{equation}
Therefore,
\begin{equation}
t\s^\mathrm{nad,FOO}[n_L,n_R](\br\rightarrow\bR_L)\approx 0.
\end{equation}
With an analogous analysis for the core region of the right nucleus, we conclude that the approximate FOO-NAKE per particle has no features in the core regions.
However, this is not the case with the ENS treatment.
In that case, the NAKE per particle can be written as
\begin{eqnarray}
t\s^\mathrm{nad,ENS}[n_L,n_R](\br)&=&\frac{1}{2}\sum_{\sigma}t\s\unpol[2(n_{L\sigma}+n_{R\sigma})](\br)\nonumber\\
& &-\frac{1}{2}\sum_{i,\sigma}f_{iL}t\s\unpol[2n_{iL\sigma}](\br)\nonumber\\
& &-\frac{1}{2}\sum_{i,\sigma}f_{iR}t\s\unpol[2n_{iR\sigma}](\br).
\end{eqnarray}
Since $t\s[n]$ is not a linear functional of $N$ in LDA or GGA, we have
\begin{equation}
t\s\unpol[2n_{L\sigma}](\br\rightarrow\bR_L)\neq \sum_{i}f_{iL}t\s\unpol[2n_{iL\sigma}](\br\rightarrow\bR_L).
\end{equation}
The $t\s$ of the right fragment in the left region can still be ignored:
\begin{equation}
t\s\unpol[2n_{iR\sigma}](\br\rightarrow\bR_L)\approx 0.
\end{equation}
Therefore,
\begin{eqnarray}
t\s^\mathrm{nad,ENS}[n_L,n_R](\br\rightarrow\bR_L)&\approx&\frac{1}{2}\sum_{\sigma}t\s\unpol[2(n_{L\sigma})](\br\rightarrow\bR_L)\nonumber\\
& &-\frac{1}{2}\sum_{i,\sigma}f_{iL}t\s\unpol[2n_{iL\sigma}](\br\rightarrow\bR_L)\nonumber\\
&\neq& 0.
\end{eqnarray}
This analysis explains why the approximate ENS-NAKE per particle has unphysical features in the core regions, and it also explains why FOO is more accurate than ENS for stretched bonds.
For stretched bonds, the majority of the contribution to the NAKE comes from the core regions because the densities are localized there.
With FOO, the NAKE correctly vanishes because the NAKE {\em per particle} does, while this is not the case for ENS.

\section{Behavior of the non-additive kinetic energy per particle}\label{sec:behav_ep_kin}

\begin{figure}[htb]
	\includegraphics*[width=\textwidth]{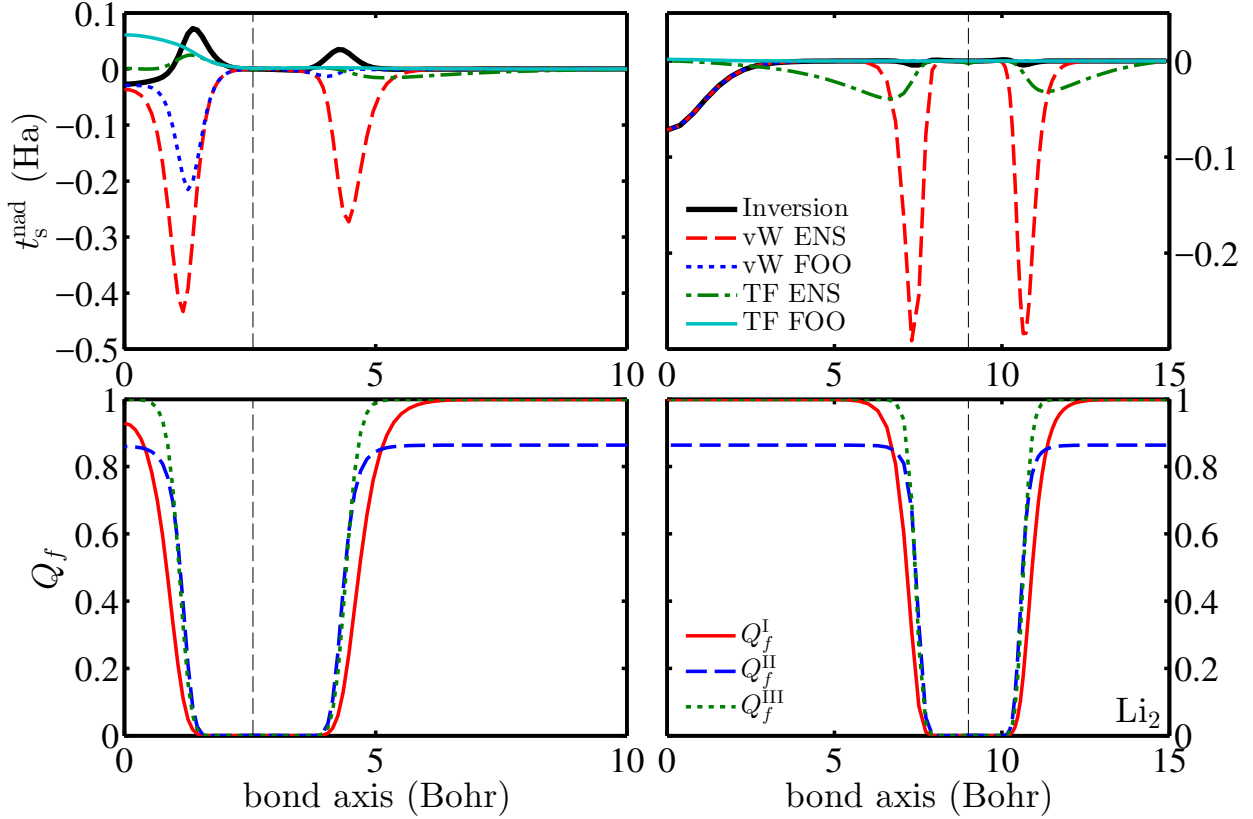} 
	\caption{Top panels: the NAKE per particle for Li$_2$. Bottom panels: Three choices of switching functions. Left panels: LDA equilibrium. Right panels: stretched configuration. The plots are evaluated along the bond axis. Only the right half is plotted due to the mirror symmetry. The black dashed straight lines indicate the positions of the (right-hand) nuclei.}
	\label{fig:Li2_epkin_Qf}
\end{figure}

\begin{figure}[htb]
	\includegraphics*[width=\textwidth]{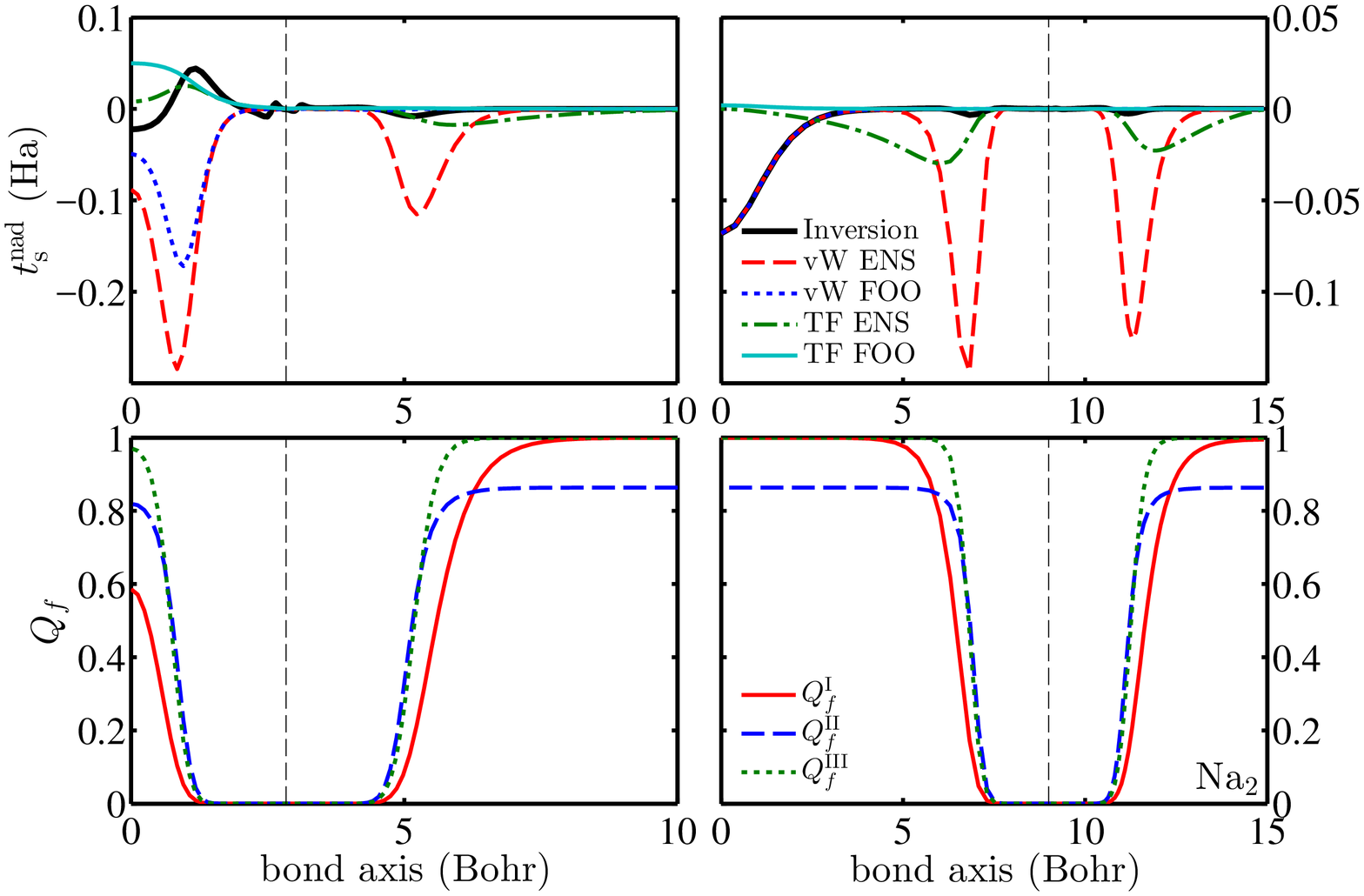} 
	\caption{Same as figure \ref{fig:Li2_epkin_Qf} with results for Na$_2$. }
	\label{fig:Na2_epkin_Qf}
\end{figure}

\begin{figure}[htb]
	\includegraphics*[width=\textwidth]{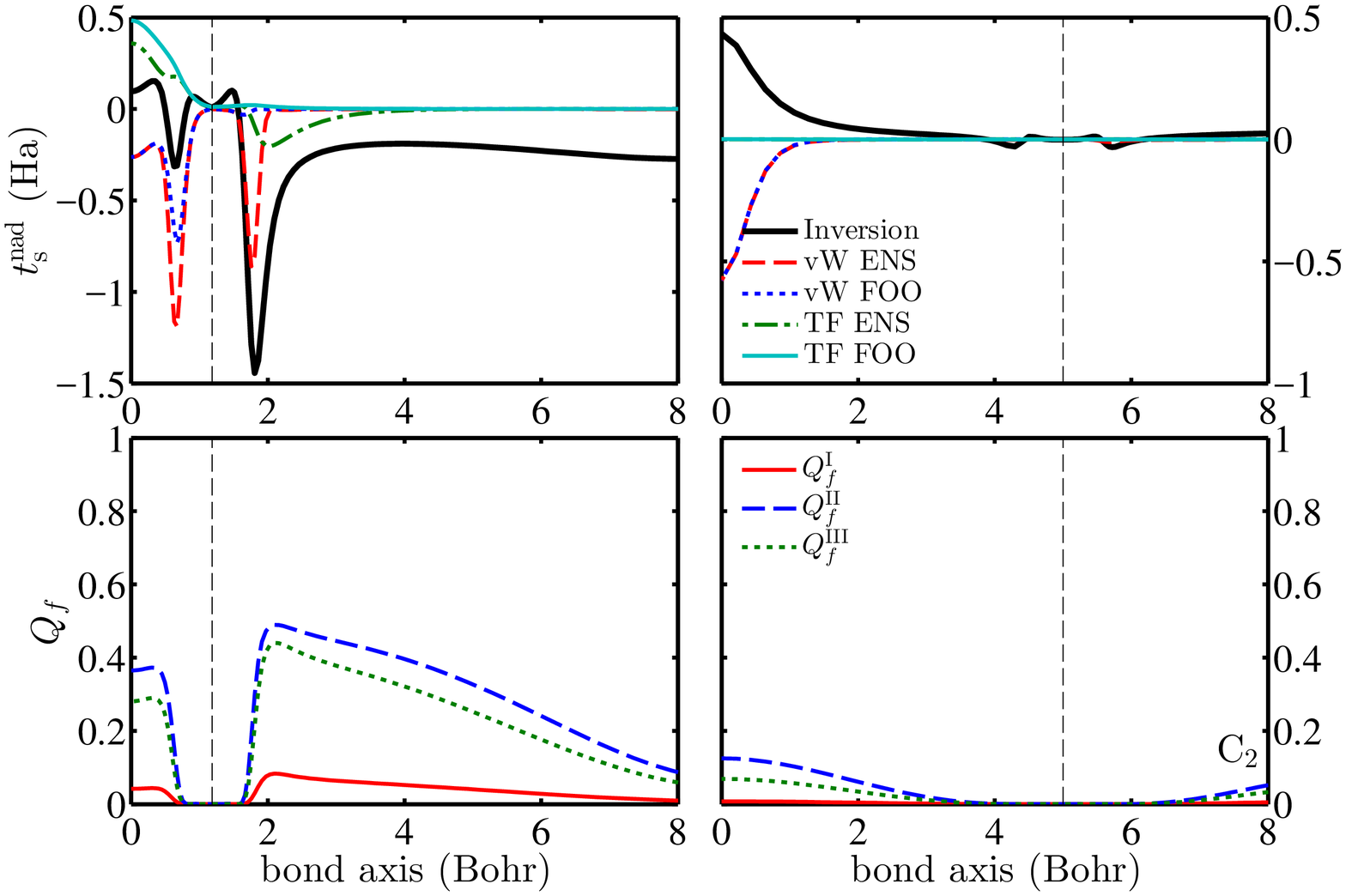} 
	\caption{Same as figure \ref{fig:Li2_epkin_Qf} with results for C$_2$.}
	\label{fig:C2_epkin_Qf}
\end{figure}

\begin{figure}[htb]
	\includegraphics*[width=\textwidth]{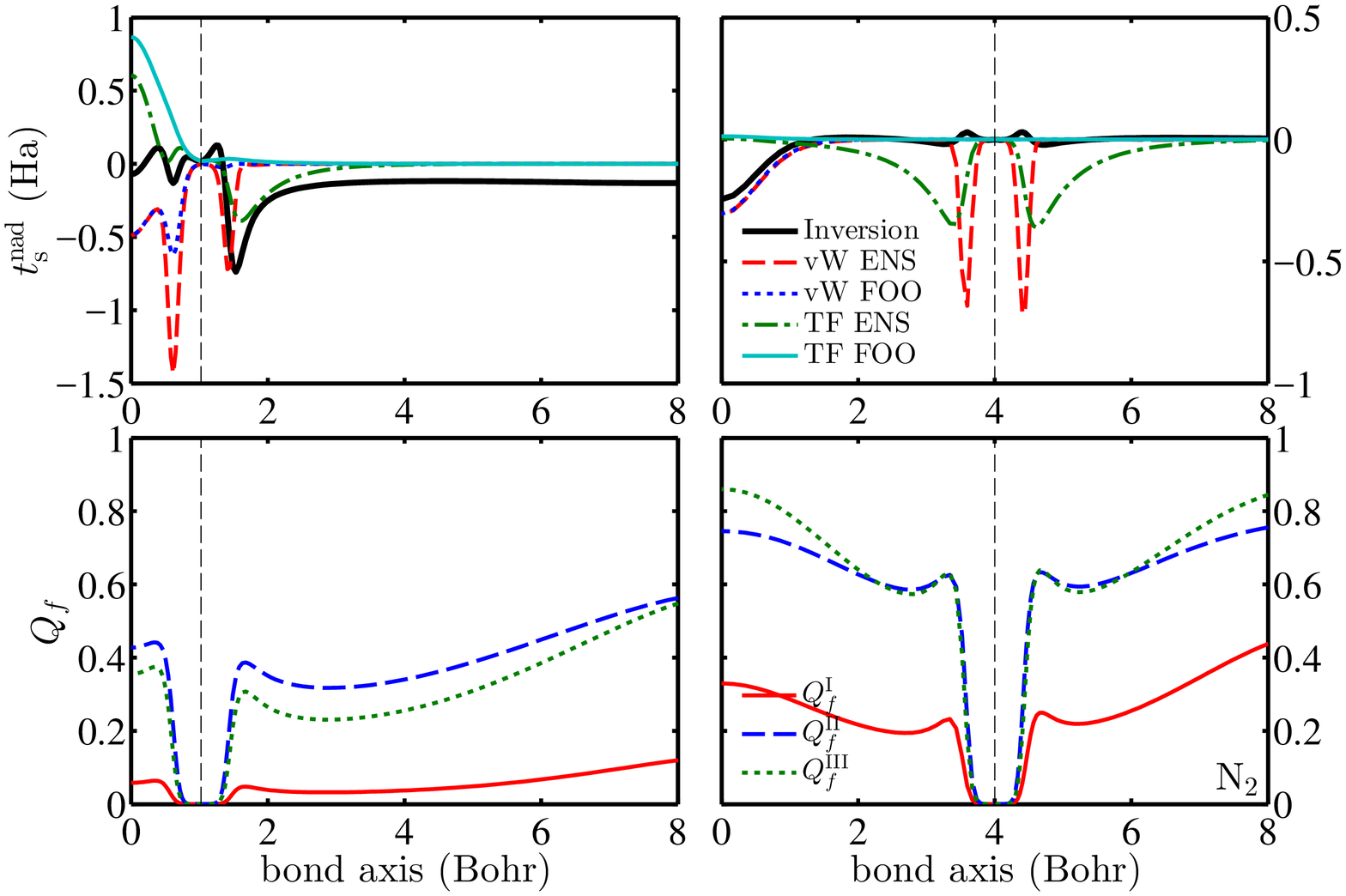} 
	\caption{Same as figure \ref{fig:Li2_epkin_Qf} with results for N$_2$.}
	\label{fig:N2_epkin_Qf}
\end{figure}

\begin{figure}[htb]
	\includegraphics*[width=\textwidth]{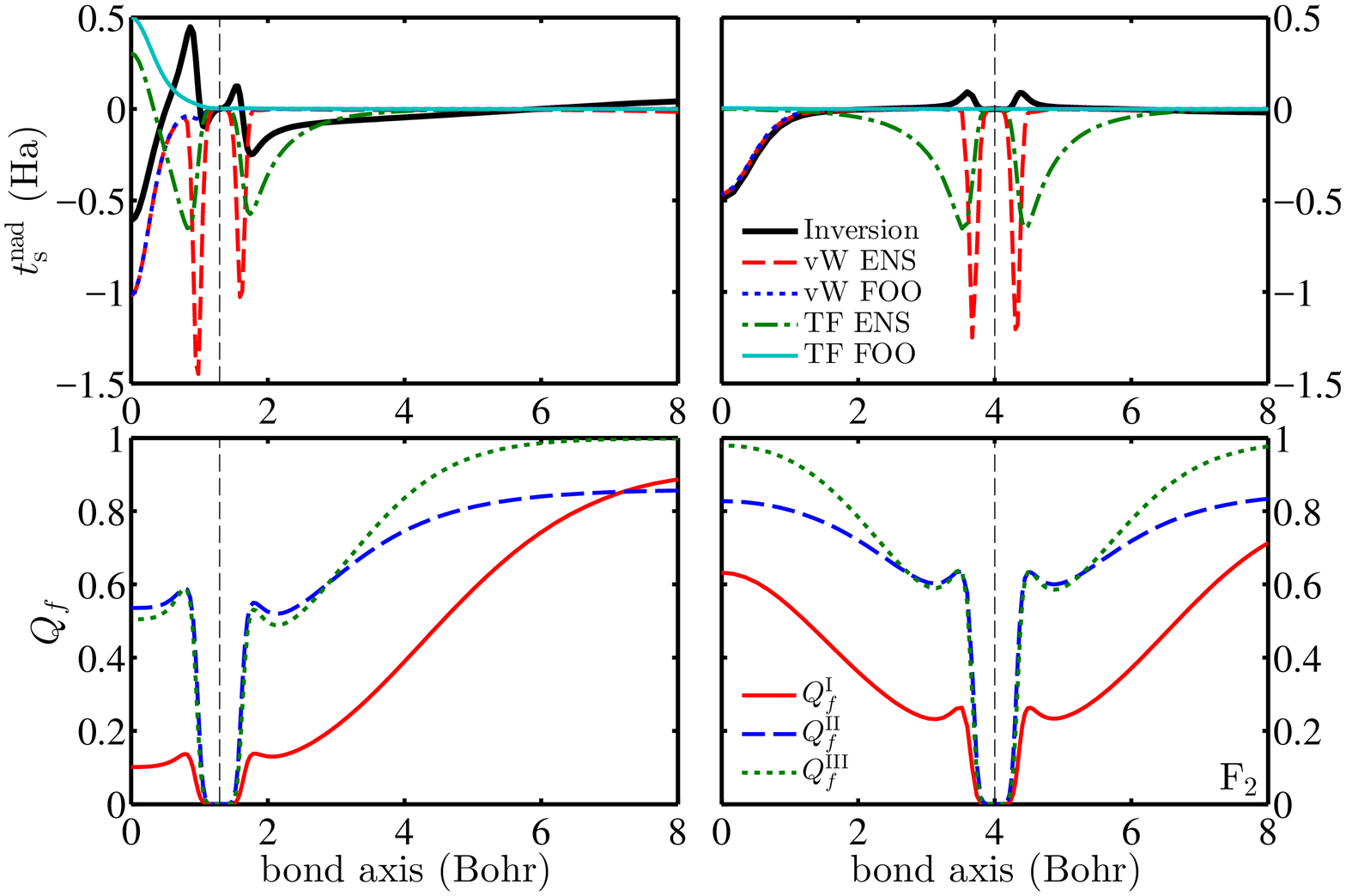} 
	\caption{Same as figure \ref{fig:Li2_epkin_Qf} with results for F$_2$.}
	\label{fig:F2_epkin_Qf}
\end{figure}

The top panels of Figures \ref{fig:Li2_epkin_Qf}-\ref{fig:F2_epkin_Qf} show the exact NAKE per particle obtained self-consistently with the inversion method and the ENS treatment for Li$_2$, Na$_2$, N$_2$, C$_2$, and F$_2$, as well as the TF and vW NAKE per particle with both ENS and FOO treatments evaluated non-self-consistently using the same fragment densities. 
With FOO, the densities are converted using Eq. \ref{eqn:density_conv}.

The exact NAKE per particle has significant features in both binding region and core region for the equilibrium configuration, while for the stretched configuration, the features in the core region are much less significant than those in the binding region.
The TF and vW NAKE per particle evaluated using the ENS treatment has features in both binding and core regions, and these features are significant in both equilibrium and stretched configurations. 
On the other hand, the TF and vW NAKE per particle evaluated with the FOO treatment has features only in the binding region, as explained in Sec \ref{sec:Tsnad}.

In stretched Li$_2$ and Na$_2$, the vW NAKE per particle is extremely close to the exact one in all but the core region with both ENS and FOO treatments, as there is only one valence orbital for alkali dimers. 
The similarity of the vW NAKE per particle and the exact NAKE per particle in the non-core region is also observed in some non-alkali dimers like N$_2$ and F$_2$.
However, in C$_2$, the exact NAKE per particle has a prominent peak in the binding region which is not reproduced by vW.

\section{Covalent approximation}\label{sec:CA}

The proposed covalent approximation (CA) is a non-decomposable NAKE constructed based on the observed similarities between the vW and the exact NAKE per particle in the non-core regions for alkali dimers.
The idea is to use the vW functional only in the regions dominated by one orbital and to use TF, the simplest NAKE functional, in all other regions:
\begin{eqnarray}\label{eqn:CA}
t\s^{\rm{nad,Covalent}}[\{n_\alpha\}](\br)&=&\Qf t\s^{\rm{nad,vW,FOO}}[\{n_\alpha\}](\br)\nonumber\\
& &+(1-\Qf)t\s^{\rm{nad,TF,FOO}}[\{n_\alpha\}](\br)
\end{eqnarray}

There have been a few attempts made to define the regions that the switching function $Q_f(\br)$ is meant to distinguish.
Sun \textit{et al.} use parameters calculated from the exact non-interacting kinetic energy density and the non-interacting kinetic energy density approximated by TF and vW. \cite{SXF+13} Lastra \textit{et al.}, on the other hand, use a function of the fragment densities, their gradient and Laplacian. \cite{LKW08}. 
To seamlessly connect the vW and TF NAKE per particle, we interpret $Q_f$ in Eq. \ref{eqn:CA} as a function of the fragment densities, 
$Q_f(\{n_{\alpha}(\br)\})$, which is also a function of space and should ideally meet the following exact constraints:
 \begin{itemize}
 	\item $0\leqslant\Qf\leqslant 1$.
 	\item $\Qf=1$ in one orbital limit.
 	\item $\Qf=0$ in uniform-gas limit.
 \end{itemize}
However, to meet the uniform-gas limit requirement $\Qf$ must be a function of $\nabla n(\br)$. 
As the CA is constructed to represent the NAKE for covalent bonds that are dominated by one orbital and the densities do not vary slowly, we believe it is more important to have $\Qf$ satisfy the one-orbital limit rather than the uniform-gas constraint.
 
We tested the following three choices for $Q_f(\{n_{\alpha}(\br)\})$:
\begin{equation}\label{eqn:Qf1}
Q_f^{\mathrm{I}}(\br)=\prod_{\alpha=1}^{N\f}\left(1+\sum_{i}f_{i\alpha}\sum_{\sigma}\dfrac{n_{i\alpha\sigma}(\br)}{n_{i\alpha}(\br)}\log_2\dfrac{n_{i\alpha\sigma}(\br)}{n_{i\alpha}(\br)}\right)
\end{equation} 
\begin{equation}\label{eqn:Qf2}
Q_f^{\mathrm{II}}(\br)=\prod_{\alpha=1}^{N\f}\left(1-\cosh^{-2}(\frac{2m_\alpha}{n_\alpha})\right)
\end{equation} 
\begin{equation}\label{eqn:Qf3}
Q_f^{\mathrm{III}}(\br)=\prod_{\alpha=1}^{N\f}\left(\frac{1}{2}-\frac{1}{2}\cos(\frac{\pi m_\alpha}{n_\alpha})\right)
\end{equation}
where $m_\alpha(\br)=\sum_{i}f_{i\alpha}|n_{i\alpha\uparrow}(\br)-n_{i\alpha\downarrow}(\br)|$.
These choices of $\Qf$ are constructed based on the fact that with the ENS treatment, in the core region the spin up and spin down fragment densities are almost the same, while in the binding region, the spin fragment densities of the two spin components are different with $m_\alpha(\br)/{n_\alpha}(\br)=1$ in the one-orbital limit.



The bottom panels of Figures \ref{fig:Li2_epkin_Qf}-\ref{fig:F2_epkin_Qf} show the three choices of the switching function for Li$_2$, Na$_2$, N$_2$, C$_2$, and F$_2$.
All three choices of the switching function go to $0$ in core regions where the vW NAKE per particle does not match the exact one. 
For Li$_2$ and Na$_2$, $Q_f^{\mathrm{I}}$ and $Q_f^{\mathrm{III}}$ approach $1$ in non-core regions as these regions are dominated by only one orbital. 
For N$_2$, C$_2$, and F$_2$, $\Qf$ varies between $0$ to $1$ in non-core regions and $Q_f^{\mathrm{I}}$ is significantly smaller than $Q_f^{\mathrm{II}}$ and $Q_f^{\mathrm{III}}$.

As indicated in Eq. \ref{eqn:CA}, the vW and TF NAKE per particle are evaluated with FOO rather than ENS. However, except for this use of FOO, all calculations are done with the ENS treatment of fractional charges and spins, as in previous P-DFT calculations \cite{NW14,NW15}.  
The resulting NAKEs are not only almost exact for stretched bonds, but they are also more accurate around equilibrium separations, as shown next.


\section{Results}\label{sec:results}

\subsection{NAKE {\em vs.} bond length}


\begin{figure}[htb]
	\includegraphics*[width=\textwidth]{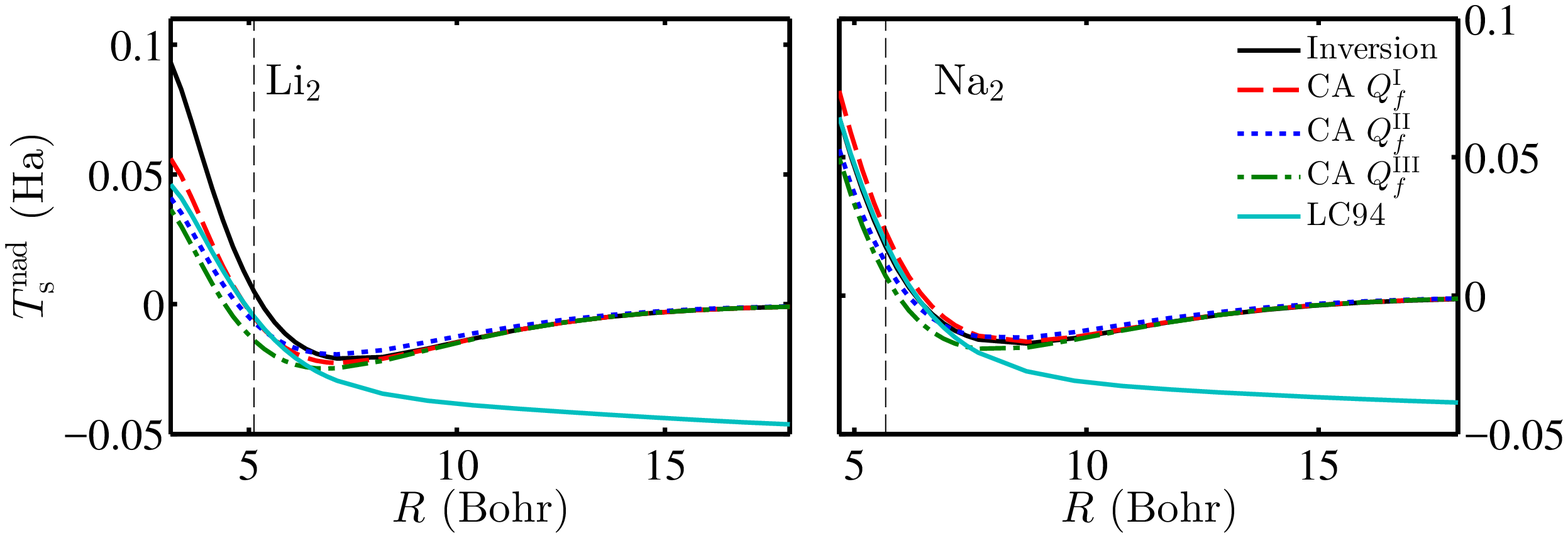} 
	\caption{NAKE {\em vs.} bond length $R$ for Li$_2$ and Na$_2$. The black dashed straight lines indicate the equilibrium separation.}
	\label{fig:Li2_Na2_Ep_kin_nsc}
\end{figure}

\begin{figure}[htb]
	\includegraphics*[width=\textwidth]{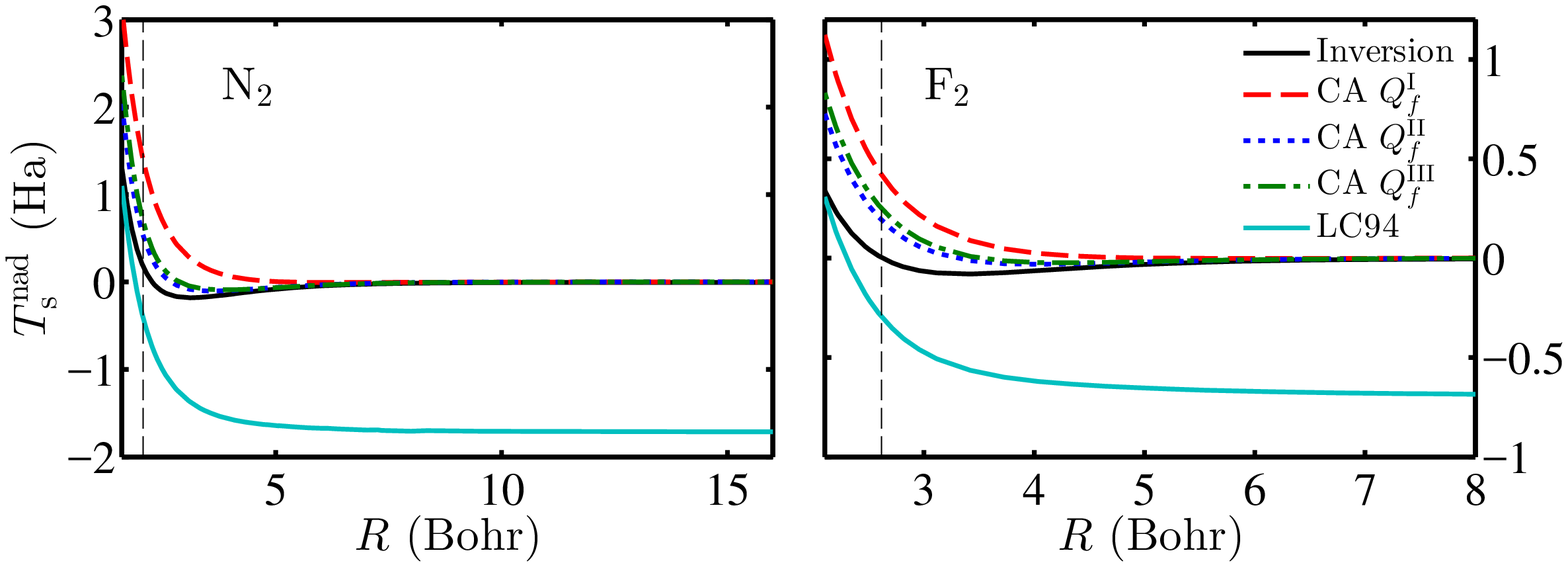}
	\includegraphics*[width=0.5\textwidth]{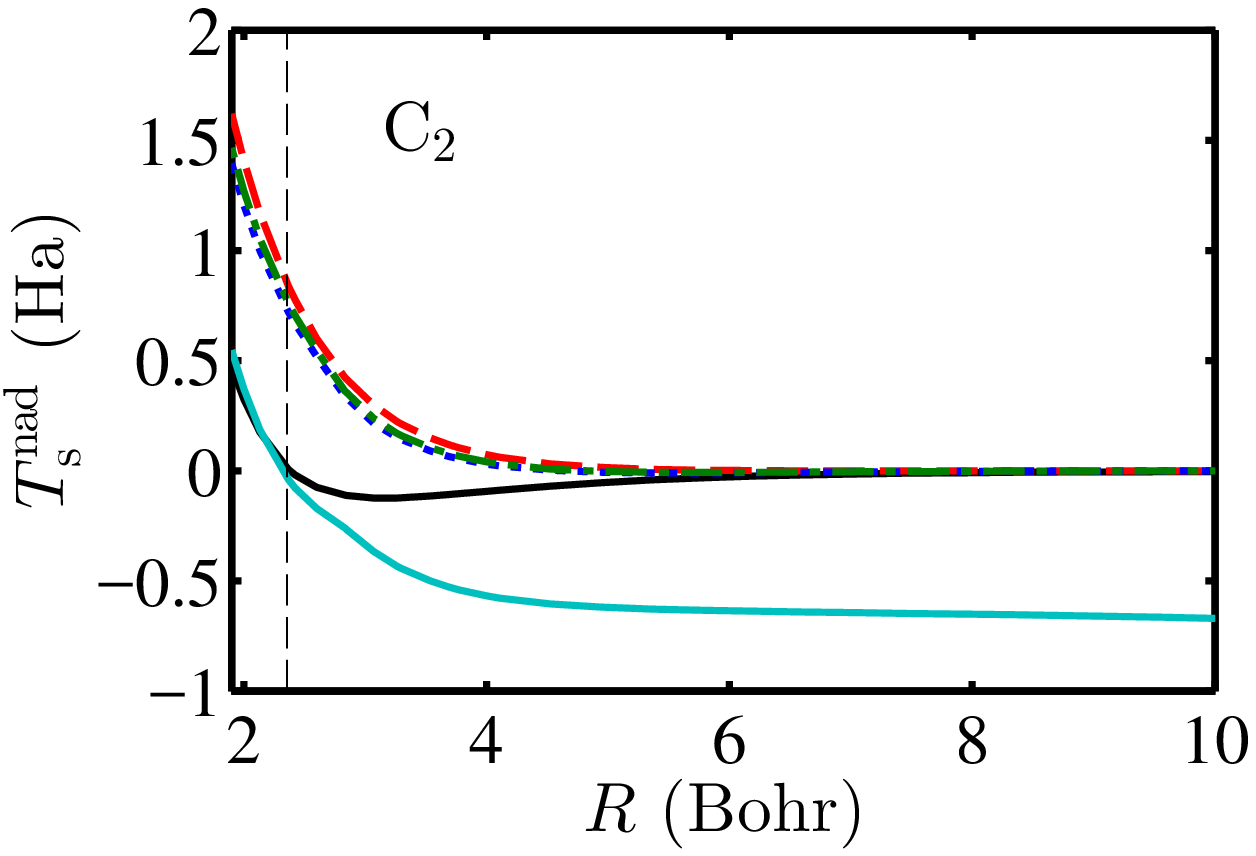} 
	\caption{NAKE vs bond length $R$ for N$_2$, F$_2$, and C$_2$. The black dashed straight lines represent the equilibrium separation.}
	\label{fig:N2_F2_Ep_kin_nsc}
\end{figure}

Figure \ref{fig:Li2_Na2_Ep_kin_nsc} displays the NAKE versus bond length for Li$_2$ and Na$_2$. 
The approximated NAKEs are calculated non-self-consistently using the density obtained from inversion (following the inversion algorithm described in ref. \cite{NJW17}).
LC94 calculated with the ENS treatment is included for comparison with existing decomposable approximations.
Our CA with all three choices of the switching function yield very accurate NAKE for stretched bonds beyond the minimum of the NAKE. 
At shorter bond length near the equilibrium, $Q_f^{\mathrm{I}}$ yields more accurate NAKE than the other two choices of the switching function, and also outperforms LC94. 

Figure \ref{fig:N2_F2_Ep_kin_nsc} displays the same data for N$_2$, F$_2$, and C$_2$.
The CA still yields very accurate NAKE for stretched bonds, but begins to diverge at a bond length farther away from the equilibrium compared to Li$_2$ and Na$_2$.
This behavior is expected in dimers with more than one valence orbital.
For N$_2$ and F$_2$, $Q_f^{\mathrm{II}}$ and $Q_f^{\mathrm{III}}$ significantly outperform $Q_f^{\mathrm{I}}$ while $Q_f^{\mathrm{II}}$ is slightly more accurate than $Q_f^{\mathrm{III}}$. 
For C$_2$, all three choices of switching function fail near the equilibrium bond length.




\subsection{Binding curves}

\begin{figure}[htb]
	\includegraphics*[width=\textwidth]{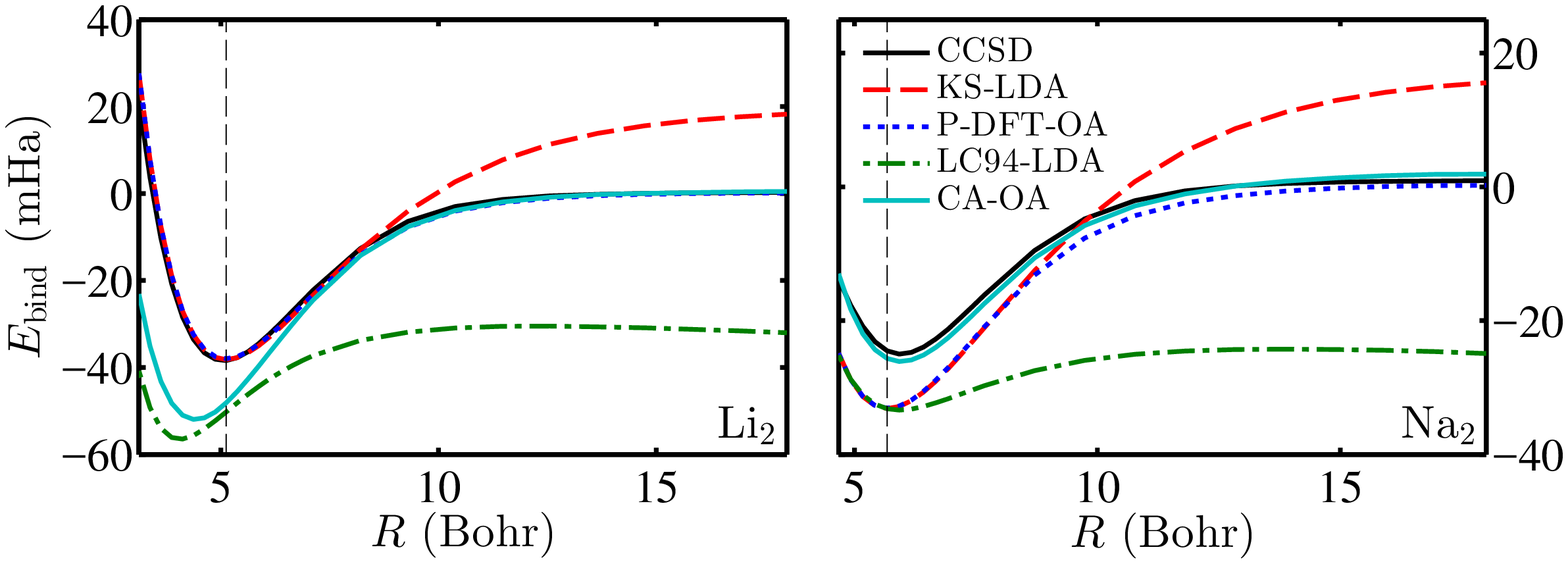} 
	\caption{Binding curves for Li$_2$ and Na$_2$. CCSD: calculated with NWChem \cite{VBG+10} with the aug-cc-pVDZ basis set used for Li$_2$ and 6-31G* used for Na$_2$. KS-LDA: KS-DFT with LDA as XC functional. P-DFT-OA: P-DFT with LDA XC functional plus corrections from the OA, and the NAKE is evaluated with inversion. LC94-LDA: P-DFT with LC94 as NAKE and LDA as XC. CA-OA: P-DFT with the CA with $Q_f^{\mathrm{I}}$ as NAKE and LDA plus the OA correction as XC. The black dashed straight lines represent the equilibrium separation.}
	\label{fig:Li2_Na2_E_bind}
\end{figure}

In this section, we demonstrate the potential of combining our CA with the overlap approximation (OA) introduced recently\cite{NW15} to treat delocalization and static-correlation errors. 
The OA is designed to maintain the KS behavior near the equilibrium while suppressing delocalization and static-correlation errors at large separations.
By combining the CA with the OA, the molecular binding energy is obtained without having to solve the KS equations either directly or inversely for the molecule as a whole, but only directly for the fragments.

Figure \ref{fig:Li2_Na2_E_bind} shows the binding curves of Li$_2$ and Na$_2$.
All calculations are done self-consistently. 
We use CCSD results as accurate reference data. \cite{VBR+83,MBC85}
P-DFT with the OA of ref.\cite{NW15} and accurate NAKE from inversion yields accurate binding energies for stretched bonds while keeping the KS-DFT description near equilibrium. 
P-DFT combining the CA for the NAKE with the OA for the non-additive XC functional also yields similarly accurate binding energy at large inter-nuclear separations. Interestingly, these two approximations act together to yield extremely accurate energies for Na$_2$ for all internuclear separations, including near equilibrium where the LDA description is poor. 

We stress that the LDA is the only approximation used here for the fragments, while the non-additive pieces were approximated through the CA of Eq. \ref{eqn:CA} for the kinetic component and the OA of ref.\cite{NW15} for the exchange-correlation component of the partition energy.

\section{Concluding remark}

Although general, robust approximations for the NAKE functional remain a challenge for covalent bonds, we have proposed the CA of Eq. \ref{eqn:CA} by examining the exact behavior of the vW NAKE per particle in regions dominated by one orbital. 
The CA connects the vW and TF with a switching function of the fragment densities that distinguishes different types of regions.
We tested three expressions for this switching function satisfying two out of the three known exact requirements.
Our CA also combines two well-known procedures for dealing with fractional spins: ENS and FOO, where the FOO form of TF and vW NAKE per particle is used to remove the static-correlation error, while the ENS method is used for calculating the switching function and fragment densities.  
We tested this CA on a few covalent dimers and observed that it reproduces the exact NAKE for stretched bonds in all systems studied here. We also observed that it outperforms existing decomposable approximations near equilibrium separations for alkali dimers.
We demonstrated the potential of the proposed approach to suppress delocalization and static-correlation errors of approximate XC functionals through P-DFT calculations that bypass the need for direct (or inverse) molecular KS calculations. 

Future work includes improving the behavior of the covalent approximation near equilibrium bond lengths.
One possibility is enforcing the uniform-gas limit to the switching function. Another obvious direction is replacing TF with a more sophisticated approximation for describing the core regions around nuclei.

\section{Acknowledgements}
We acknowledge support from the National Science Foundation CAREER program under Grant No. CHE-1149968.

\bibliography{JNW17b}

\end{document}